\documentclass[12pt,preprint]{emulateapj}

\usepackage{amsmath}
\usepackage{stmaryrd}
\usepackage{natbib}
\usepackage{hyperref}
\hypersetup{colorlinks=true,linkcolor=black,citecolor=blue,urlcolor=black}

\begin{document}

\title{Effects of Radius and Gravity on the Inner Edge of the Habitable Zone}

\author{Huanzhou Yang$^{1}$, Thaddeus D. Komacek$^{2}$, Dorian S. Abbot$^{2}$} \affil{$^1$Department of Atmospheric and Oceanic Sciences, School of Physics, Peking University, Beijing, 100871, China \\ $^2$Department of the Geophysical Sciences, The University of Chicago, Chicago, IL, 60637 \\
\url{jeff\_yang2015@pku.edu.cn}} 
\begin{abstract}
A rigorous definition of the habitable zone and its dependence on planetary properties is part of the search for habitable exoplanets. In this work, we use the general circulation model {\tt ExoCAM} to determine how the inner edge of the habitable zone of tidally locked planets orbiting M dwarf stars depends on planetary radius, surface gravity, and surface pressure. We find that the inner edge of the habitable zone for more massive planets occurs at higher stellar irradiation, as found in previous one-dimensional simulations. We also determine the relative effects of varying planetary radius and surface gravity. Increasing the planetary radius leads to a lower planetary albedo and warmer climate, pushing the inner edge of the habitable zone to lower stellar irradiation. This results from a change in circulation regime that leads to the disruption of the thick, reflective cloud deck around the substellar point. Increasing gravity increases the outgoing longwave radiation, which moves the inner edge of the habitable zone to higher stellar irradiation. This is because the column mass of water vapor decreases with increasing gravity, leading to a reduction in the greenhouse effect. The effect of gravity on the outgoing longwave radiation is stronger than the effect of radius on the planetary albedo, so that increasing gravity and radius together causes the inner edge of the habitable zone to move to higher stellar irradiation. Our results show that the inner edge of the habitable zone for more massive terrestrial planets occurs at a larger stellar irradiation. 
\end{abstract}

\section{Introduction}
The search for extant life on exoplanets with future space telescopes such as \textit{JWST}, \textit{LUVOIR}, \textit{HabEx} and \textit{OST} requires narrowing the phase space of potentially inhabited exoplanets. One criterion often used for habitability is the presence of surface liquid water, which is necessary for Earth-like life to flourish. The criterion of surface liquid water for habitability has led to the development of the habitable zone concept \citep{Kasting:1993aa}, along with a variety of recent work establishing the limits of the habitable zone for Earth-like planets around Sun-like stars \citep{Kopparapu:2013aa,Leconte:2013,Kopparapu:2014,Wolf:2015,Popp:2016,Way:2016,Wolf:2017} and lower-mass M dwarfs \citep{Yang:2013,Kopp:2016,kopparapu2017,Yang:2019aa}. The maximum stellar irradiation at which a planet is habitable (the ``inner edge'' of the habitable zone) is expected to be larger for planets orbiting M dwarf stars than for planets orbiting Sun-like stars. This is because planets orbiting M dwarf stars are tidally locked to their host stars. The resulting strong irradiation on the dayside leads to convection which causes extensive cloud cover. This cloud cover reduces the absorbed stellar irradiation and cools the surface of the planet \citep{Yang:2013,Yang:2014,kopparapu2017,Genio:2017aa,way:2018,Komacek:2019aa}. \\
\indent In this work, we focus on how varying planetary radius and surface gravity affect the inner edge of the habitable zone for tidally locked planets orbiting M dwarf stars. The effect of varying planetary mass on the inner edge of the habitable zone has previously been explored using a 1D radiative-convective, cloud-free climate model by \cite{Kopparapu:2014}. \citeauthor{Kopparapu:2014} used scaling relations between the planetary radius, surface gravity, surface pressure, and planetary mass to determine how the inner edge of the habitable zone depends on planetary mass. \citeauthor{Kopparapu:2014} found that the inner edge of the habitable zone moves to higher stellar irradiation with increasing planetary mass due to the reduced water column optical depth for more massive planets. This is because more massive planets have correspondingly larger surface gravities, so that less water vapor mass is required to produce the same vapor pressure \citep{Kopparapu:2014,Thomson:2019aa}. This reduces the greenhouse effect, which increases the limiting value of outgoing longwave radiation that is reached in the runway greenhouse state. \\
\indent Here we reinvestigate the effect of planet mass on the inner edge of the habitable zone using the 3D GCM {\tt ExoCAM}. We will vary planetary radius and surface gravity separately, as well as planetary mass as in \cite{Kopparapu:2014}. We consider a range of planetary radii from $0.5-2~R_\varoplus$ and surface gravities from $0.4-1.6~g_\varoplus$ to encompass a wide range of exoplanets, from Mars-sized planets to super-Earths. The outline of this paper is as follows. In Section \ref{sec:Methods} we describe the {\tt ExoCAM} GCM and our experimental setup. In Section \ref{sec:Res}, we display our results for how the inner edge of the habitable zone depends on planetary radius, gravity, and surface pressure. We discuss in detail the physical mechanisms by which varying radius and gravity separately affect the inner edge of the habitable zone in Section \ref{sec:Dis}, and state our conclusions in Section \ref{sec:conc}.
\section{Methods} \label{sec:Methods}
In this work, we use the {\tt ExoCAM} GCM\footnote{\url{https://github.com/storyofthewolf/ExoCAM}}, an adapted version of the Community Atmosphere Model version 4 with a novel correlated-k radiative transfer scheme and updated water vapor absorption coefficients. {\tt ExoCAM} has previously been used for a wide variety of exoplanet studies \citep{Wolf:2015,Kopp:2016,kopparapu2017,Wolf:2017,Wolf:2017aa,Haqq2018,Komacek:2019aa}. We use the same setup of {\tt ExoCAM} as \cite{kopparapu2017,Haqq2018,Komacek:2019aa}, consisting of an aquaplanet (no continents) with a slab (immobile) ocean of $50~\mathrm{m}$ depth and an atmosphere comprised purely of N$_2$ and H$_2$O. The planets we simulate orbit M dwarf stars with an effective temperature of $3700~\mathrm{K}$, with stellar spectra taken from the BT-SETTL models of \cite{Allard:2007aa}. We use a horizontal resolution of $4^\circ \times 5^\circ$ with 40 vertical levels, as in \cite{Komacek:2019aa}. Note that we do not consider continents, which have been found to affect the dayside cloud coverage of M dwarf planets \citep{Lewis:2018aa}. We also do not consider ocean dynamics, which \cite{Yang:2019aa} found do not strongly affect the inner edge of the habitable zone for tidally locked planets. \\
\indent We computed four types of simulations with our GCM, as shown in Table 1: (1) varying radius alone; (2) varying gravity alone; (3) varying radius and gravity together; (4) varying radius, gravity, and surface pressure together. For our simulations varying planetary radius alone, we considered radii of $0.5,~1,~2~R_\varoplus$, while keeping the gravity fixed to that of Earth. For our simulations varying surface gravity alone, we considered gravities of $0.4,~0.8,~1.6~g_\varoplus$, keeping the radius fixed to that of Earth. When varying planetary radius and gravity together, we ran one model with a radius and gravity equal to that of Earth and ran a second model using the scaling relations of \cite{Kopparapu:2014} to calculate the gravity and planetary mass assuming a radius of $0.5~R_\varoplus$. We used the same radius and gravity when varying radius, gravity, and pressure together, but also calculated the surface pressure using the scaling relation of \cite{Kopparapu:2014}. \\
\indent To find the inner edge of the habitable zone for a given combination of planetary parameters, we started with two simulations -- one simulation at high enough stellar irradiation to be in a runaway state, and one cool enough to be stable. We consider simulations with yearly-average net radiative imbalance of less than $2~\mathrm{W}~\mathrm{m}^{-2}$ to be stable, and simulations that continue to warm and crash due to extreme high temperature to be in a runaway state. Some of the simulations stayed in a fluctuating climate state for more than 100 years. We consider these simulations to be stable, because the average temperature stays similar over long timescales. \\
\indent We used a bisection algorithm to find the inner edge of the habitable zone to within $\pm 2.5~\mathrm{W}~\mathrm{m}^{-2}$ in stellar irradiation. In all of our simulations, we varied the rotational period self-consistently with the stellar irradiation using Equation (1) of \cite{kopparapu2017}, assuming that the rotation period is equal to the orbital period. We performed numerical tests of our method for identifying the inner edge of the habitable zone with varying timesteps of $30,~15,~\mathrm{and}~7.5~\mathrm{min}$, finding that the inner edge does not qualitatively change for timesteps less than $15~\mathrm{min}$. As a result, we use a dynamical timestep of $15~\mathrm{min}$ in this work, with radiative transfer computed every three dynamical timesteps.\\
\indent To better analyze our simulations of tidally locked planets, we utilize the tidally locked coordinate system \citep{Koll:2014} in this work. In this coordinate system, the ``tidally locked'' (TL) north pole and south pole are set at the substellar point and the antistellar point, respectively. The new equator is the border between the day and night hemispheres (the terminator). The climates on slowly rotating tidally locked planets with a rotation period of about 30-40 days exhibit a strong symmetry about the axis connecting the substellar and antistellar points. As a result, the tidally locked coordinate system allows us to calculate physically meaningful zonal averages, while calculating zonal mean physical quantities in standard coordinates does not provide physically meaningful information. We refer the reader to Appendix B of \cite{Koll:2014} for the mathematical translation between standard and tidally locked coordinates\footnote{The code we use to translate GCM output to tidally locked coordinates can be found at: \url{https://github.com/ddbkoll/tidally locked-coordinates}}.

\section{The effect of varying planetary size on the inner edge of the habitable zone} \label{sec:Res}
\begin{deluxetable*}{ccccccc}[h!] \label{tab:mathmode}
\centering
\tablecaption{
Summary of the numerical experiments. The inner edge of the habitable zone in stellar irradiation and the corresponding orbital period at the inner edge are shown in the two rightmost columns for each of our sets of assumed planetary parameters. The leftmost column groups our experiments into those varying radius only; gravity only; radius and gravity together; and varying radius, gravity, and pressure together.}
\tablecolumns{7}
\tablenum{1}
\tablewidth{0pt}
\tablehead{
\colhead{Experiment Group} &
\colhead{$R/R_{\varoplus}$} &
\colhead{$g/g_{\varoplus}$} &
\colhead{$M/M_{\varoplus}$} &
\colhead{$p_s/p_{s,\varoplus}$} & 
\colhead{Inner Edge ($\mathrm{W}~\mathrm{m}^{-2}$)} &
\colhead{Orbital Period (Earth Days)}
}
\startdata
Radius only & 0.5 & 1 & 0.25 & 1 & 1902.5 & 37.11\\
& 1 & 1 & 1 & 1 & 1857.5 & 37.78\\
& 2 & 1 & 4 & 1 & 1777.5 & 39.05\\
\hline \\
Gravity only & 1 & 0.4 & 0.4 & 1 & 1692.5 & 40.52\\
& 1 & 0.8 & 0.8 & 1 & 1832.5 & 38.17\\
& 1 & 1.6 & 1.6 & 1 & 1937.5 & 36.61\\
\hline \\
Radius and gravity & 0.5000 & 0.4213 & 0.1053 & 1 & 1722.5 & 39.99\\
 & 1 & 1 & 1 & 1 & 1857.5 & 37.78\\
\hline \\
Radius, gravity, and pressure & 0.5000 & 0.4213 & 0.1053 & 0.1775 & 1627.5 & 41.72\\
& 1 & 1 & 1 & 1 & 1857.5 & 37.78
\enddata
\end{deluxetable*}
\indent Our results for the stellar irradiation at the inner edge of the habitable zone from our four different types of simulations are shown on the right hand side of Table 1. We find that increasing the planetary radius from $0.5-2~R_\varoplus$ (top three rows in Table 1) causes the inner edge to occur at a stellar irradiation that is $125~\mathrm{W}~\mathrm{m}^{-2}$ lower. Conversely, increasing the surface gravity from $0.4-1.6~g_\varoplus$ (rows 4-6 in Table 1) causes the inner edge to occur a stellar irradiation that is $245~\mathrm{W}~\mathrm{m}^{-2}$ higher. We find that the effect of surface gravity is larger than the effect of planetary radius on the inner edge of the habitable zone: when we increase the radius and gravity together (rows 7-8 in Table 1), we find that the inner edge of the habitable zone occurs at a higher stellar irradiation.
We will discuss the mechanisms causing these results in Section \ref{sec:Dis}. \\
\indent When we vary the surface pressure with radius and gravity (bottom row of Table 1), we find the same trend of increasing stellar irradiation at the inner edge with increasing planetary size, but with an even larger magnitude ($230~\mathrm{W}~\mathrm{m}^{-2}$ relative to $135~\mathrm{W}~\mathrm{m}^{-2}$ when we vary just radius and gravity). 
This was also found in the 1D simulations of \cite{Kopparapu:2014}, and is because increased surface pressure leads to enhanced Rayleigh scattering, a higher albedo, and a cooler climate. \\
\indent Because we use the same scaling relationship between planetary mass, radius, surface gravity, and surface pressure, we can compare our results for varying planetary mass with those of \cite{Kopparapu:2014}. For planets with masses of approximately $0.1~M_\varoplus$ and $1~M_\varoplus$, respectively, we find an inner edge of $1.19~S_\varoplus$ and $1.36~S_\varoplus$, while \cite{Kopparapu:2014} found an inner edge of $0.85~S_\varoplus$ and $0.93~S_\varoplus$ for the lowest-mass stellar systems they modeled, where $S_\varoplus$ is Earth's insolation. 
We therefore find that the stellar irradiation at the inner edge of the habitable zone is larger when we increase planetary mass, in agreement with \cite{Kopparapu:2014}. However, we find that the inner edge of the habitable zone occurs at a higher stellar irradiation than \cite{Kopparapu:2014}, and we find a large increase in stellar irradiation at the inner edge when we increase planetary mass. The key difference between our simulations and those of \cite{Kopparapu:2014} is that our simulations are three-dimensional, tidally locked, and include the effects of clouds. \\
\indent Table 1 shows the orbital period at the inner edge of the habitable zone from our suite of simulations, which assume an M dwarf host star with an effective temperature of $3700 \ \mathrm{K}$. We find that the inner edge from our most realistic simulations including varying radius, gravity, and pressure occurs at an orbital period of 41.72 days for $0.5R_\varoplus$ planets and an orbital period of 37.78 days for $1R_\varoplus$ planets. The effective temperatures of \textit{TESS} target stars peaks at $\sim 3400 \ \mathrm{K}$ \citep{Sullivan:2015}. As a result, we expect that the \textit{TESS} mission can find terrestrial planets near the inner edge of the habitable zone orbiting bright M dwarf stars with similar effective temperatures to the host star we consider. Follow-up secondary eclipse and/or phase curve observations of these planets with \textit{JWST} could determine whether or not our expectation that the inner edge of the habitable zone occurs at higher stellar irradiation with increasing planetary mass is correct. This could be done through determining the surface temperature of the planet at secondary eclipse or by analyzing the morphology of infrared phase curves \citep{Yang:2013,Haqq2018,Yang:2019aa}.
\begin{figure*}
\centering
\includegraphics[width=1\textwidth]{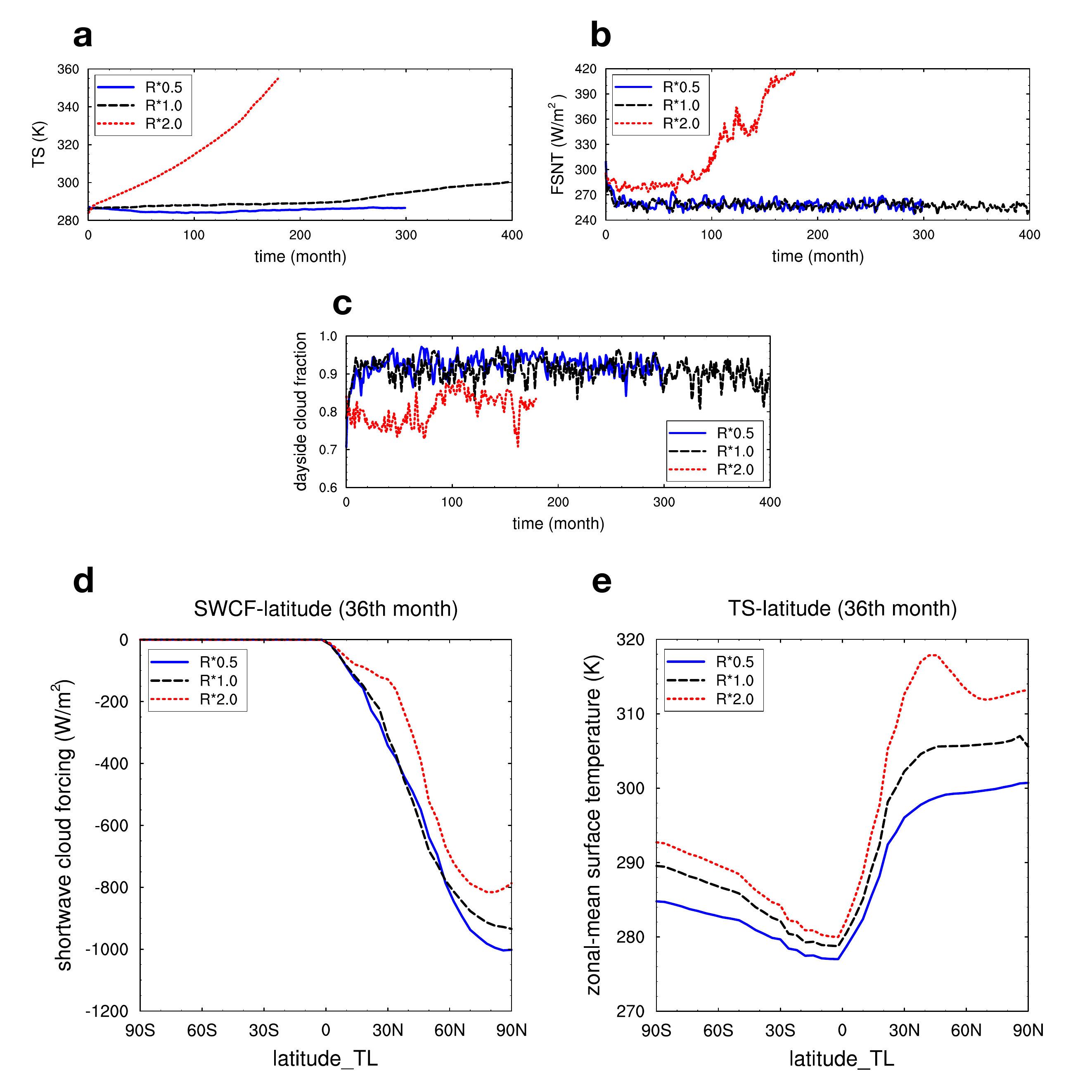}
\caption{The impact of varying planetary radius on the time series of global mean temperature (a), net shortwave flux at the top of the model (b), and the vertically integrated total cloud on the dayside (c). We also show the meridional dependence of the shortwave cloud forcing (d) and surface temperature (e) as a function of tidally locked latitude (see Appendix) from the 36th month of each simulation. All three simulations have the same stellar irradiation of $1840~\mathrm{W}~\mathrm{m}^{-2}$, which is close to the inner edge of the habitable zone for a planet with a radius equal to that of Earth.}
\end{figure*}
\begin{figure*}
\centering
\includegraphics[width=1\textwidth]{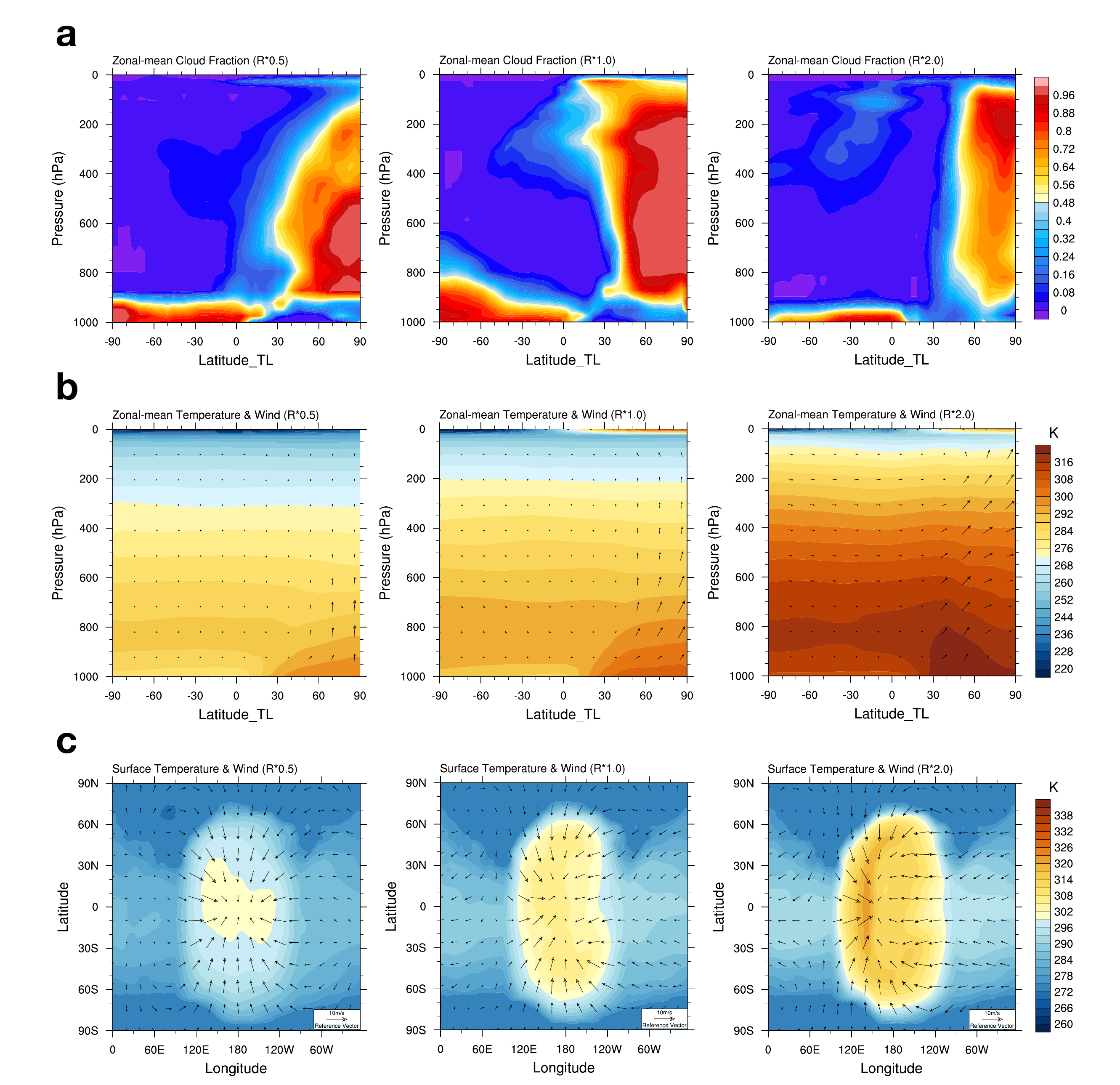}
\caption{The tidally locked zonal-mean cloud fraction (a), zonal mean temperature \& wind (b), and surface temperature and surface wind (c) in the $0.5R_\varoplus$ (left), $1.0R_\varoplus$ (middle) and $2.0R_\varoplus$ (right) simulations with a stellar irradiation of $1840~\mathrm{W}~\mathrm{m}^{-2}$.}
\end{figure*}

\section{The impact of planetary radius and surface gravity on the inner edge of the habitable zone} \label{sec:Dis}
\subsection{Varying Planetary Radius}
The inner edge of the habitable zone locates at lower stellar irradiation for planets with larger radius. To analyze the reason for this, we conducted three simulations with planetary radii of $0.5R_\varoplus$, $1.0R_\varoplus$ and $2.0R_\varoplus$, where $R_\varoplus$ is the radius of Earth. We set the stellar irradiation to 1840 W m$^{-2}$ in all three simulations, which is close to the inner edge of the $1.0R_\varoplus$ simulation, so that the three simulations result in different climate states. As illustrated in Figure 1a, the simulation with $0.5R_\varoplus$ reached a balanced state in 300 months and the simulation with $2.0R_\varoplus$ entered a runaway state. The simulation with $1.0R_\varoplus$ reached a balanced state after more than 1000 months. This analysis will focus on the early period when the simulations started to show differences in their temperature evolution. \\
\indent A reduction in albedo caused by less cloud coverage on the dayside is the cause of the increase in temperature with increasing radius. The $2.0R_\varoplus$ simulation has a much higher net shortwave flux at top of model (FSNT, see Figure 1b) than the other two simulations. This results from reduced cloud reflection (Figure 1d) due to the reduced cloud fraction on the dayside (Figure 1c) in the $2.0R_\varoplus$ simulation.

The cloud deck around the substellar point becomes thinner as the radius increases (Figure 2a). The cloud deck in the $2.0R_\varoplus$ simulation shows a large difference from the other two simulations that did not go into a runaway state. In the simulations that did not reach a runaway, a substellar cloud deck is formed because of strong atmospheric upwelling at the substellar point. We find that the upwelling in the lower troposphere becomes weaker with larger planetary radius (Figure 2b). In general, the strongest convective upwelling occurs where the surface temperature is the highest. In the $0.5R_\varoplus$ and the $1.0R_\varoplus$ simulations, the highest surface temperatures are located at the substellar point.
In the $2.0R_\varoplus$ simulation, the highest surface temperature is at a tidally locked latitude of $\approx 40^\circ$, and convergence also occurs at this latitude. Additionally, there are no longer strong updrafts near the surface at the substellar point. As a result, the cloud deck is thinner in the $2.0R_\varoplus$ simulation relative to planets with smaller radii.
\\ \indent We find that the decrease in cloud coverage with increasing planetary radius is due to a change in the dynamical state of the atmosphere. For a continuously stratified fluid, the Rossby radius of deformation, 
\begin{equation}
L_R \equiv \frac{NH}{f_0},
\end{equation}
where $N$ is the buoyancy frequency, $H$ is the scale height, and $f_0$ is the Coriolis frequency, does not change significantly with radius. When gravity is fixed, the ratio of the Rossby radius of deformation to the planetary radius, $\frac{L_R}{R} \equiv \frac{NH}{f_0R}$, decreases as planetary radius $R$ increases. This leads to a predominantly east-west flow at the substellar point instead of a symmetric climate with low-level flow converging on the substellar point from all directions \citep{Carone:2015aa,Haqq2018}. As a result, the maximum in temperature is advected westward from the substellar point (Figure 2c), so that the temperature no longer monotonically increases with tidally locked latitude (Figure 1e). This leads to the disruption of the cloud deck and an increase in global temperature, eventually causing a runaway greenhouse.
\subsection{Varying Surface Gravity}
\begin{figure}
\centering
\includegraphics[width=0.475\textwidth]{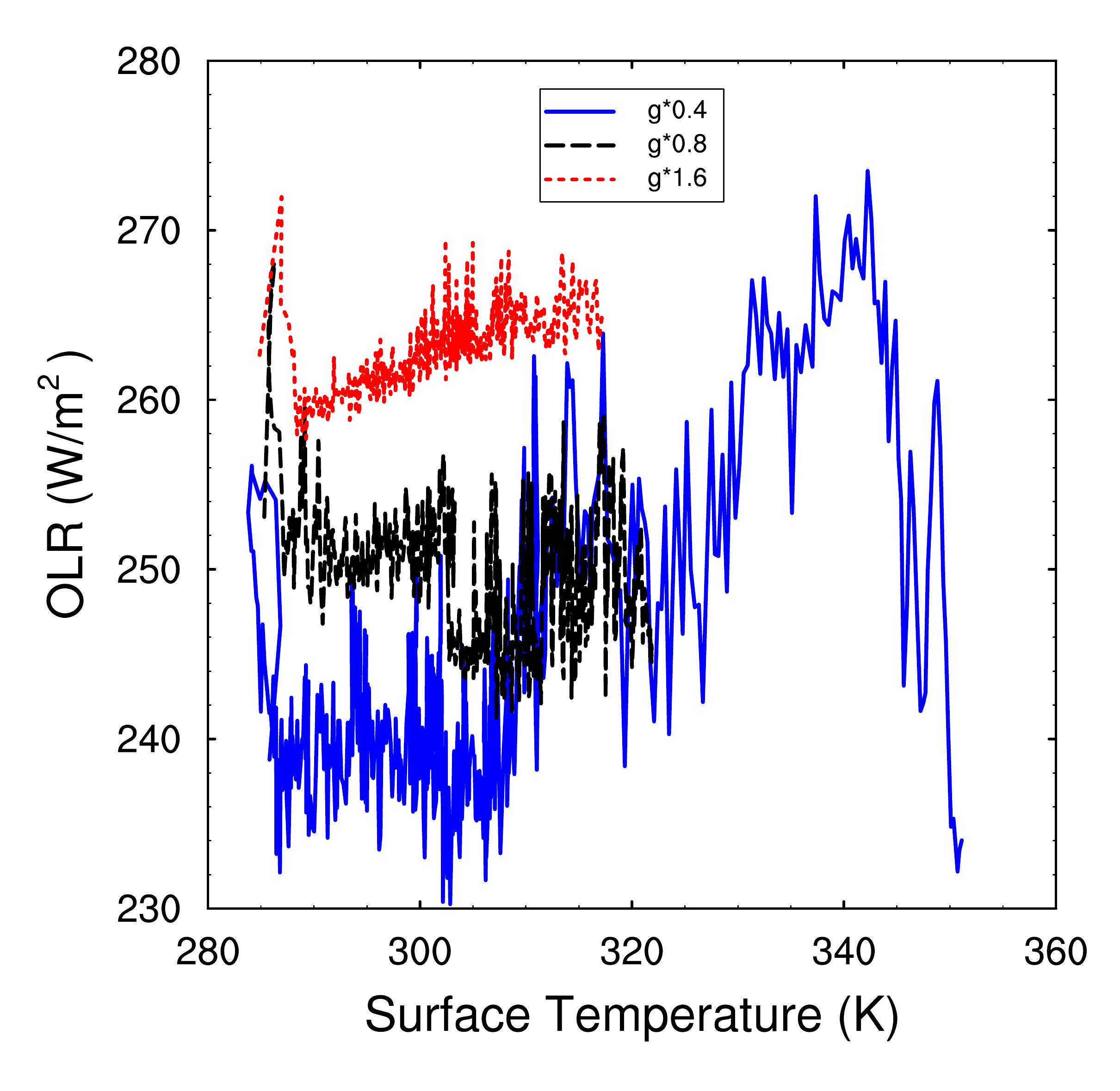}
\caption{GCM outgoing longwave fluxes with respect to global-mean surface temperature in the $0.4g_\varoplus$ simulation with a stellar irradiation of 1700 $\mathrm{W}~\mathrm{m}^{-2}$ (blue), the $0.8g_\varoplus$ simulation with a stellar irradiation of 1840 $\mathrm{W}~\mathrm{m}^{-2}$ (black), and the $1.6g_\varoplus$ simulation with a stellar irradiation of 1940 $\mathrm{W}~\mathrm{m}^{-2}$ (red).}
\end{figure}
The impact of gravity on the inner edge is opposite to that of planetary radius: larger gravity causes the inner edge of the habitable zone to move to higher stellar irradiation. The inner edge is partially determined by the maximum value of outgoing longwave flux ($\mathrm{OLR}_\infty$). The larger $\mathrm{OLR}_\infty$ is, the larger the stellar irradiation at the inner edge can be. Gravity is one of the main factors that influence $\mathrm{OLR}_\infty$. 
As discussed in \cite{Pierrehumbert:2010} and \cite{Kopparapu:2014}, $\mathrm{OLR}_\infty$ increases with gravity, because less water vapor mass is necessary to produce a given water vapor pressure. \\
\indent Figure 3 plots $\mathrm{OLR}$ from our GCM results with respect to global-mean surface temperature. 
We show simulations that all have a stellar irradiation just above the value necessary to enter a runaway state. There is a plateau in the $T-\mathrm{OLR}$ curve and the value of $\mathrm{OLR}$ at this plateau reflects $\mathrm{OLR}_\infty$. 
We find that the OLR represented by the plateau increases as the gravity increases, as expected from \cite{Pierrehumbert:2010} and \cite{Kopparapu:2014}. In the 0.4 $g_\varoplus$ simulation, the OLR eventually increases beyond this plateau when the climate enters a runaway state in which the atmosphere is hotter as well as significantly drier than saturation.


\subsection{Varying Radius and Gravity}
\begin{figure*}
\centering
\includegraphics[width=1\textwidth]{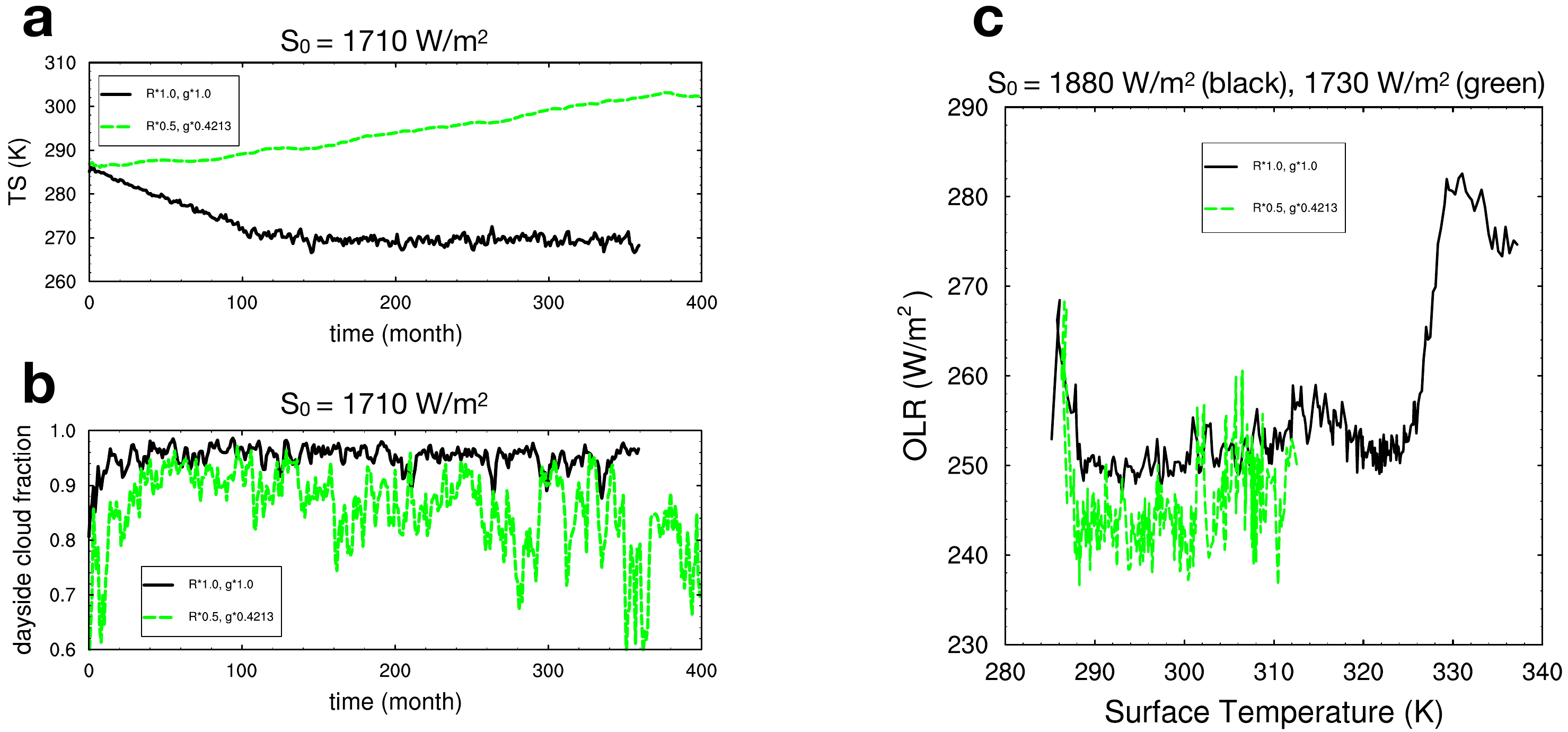}
\caption{The impact of varying planetary radius and gravity simultaneously on the time series of global mean temperature (a) and the vertically integrated total cloud on the dayside (b). Panels (a) and (b) show results from simulations with the same stellar irradiation of $1710~\mathrm{W}~\mathrm{m}^{-2}$, which is close to the inner edge of the habitable zone for the $0.5R_\varoplus, 0.4213g_\varoplus$ simulation. Panel (c) shows GCM outgoing longwave fluxes with respect to global-mean surface temperature from the $1.0R_\varoplus, 1.0g_\varoplus$ simulation with a stellar irradiation of 1880 $\mathrm{W}~\mathrm{m}^{-2}$ and the $0.5R_\varoplus, 0.4213g_\varoplus$ simulation with a stellar irradiation of 1730 $\mathrm{W}~\mathrm{m}^{-2}$, both of which are in a runway state.
}
\end{figure*}
\indent The combined effect of increasing radius and gravity causes the inner edge of the habitable zone to occur at higher stellar irradiation. There is a similar dependence as when varying gravity alone, as the effect of gravity on OLR$_\infty$ outweighs the effect of radius. Figure 4 shows the dependence of climate on varying radius and gravity together. We find that the global-mean $OLR_\infty$ is smaller for less massive planets (Figure 4c). \\
\indent As discussed in Section 4.2, because of the stronger longwave radiation absorption of water vapor, reduced gravity can lead to an increase in the global mean temperature. We find that simulations with reduced radius and gravity are significantly hotter than simulations with an Earth-like radius and gravity (Figure 4a). Additionally, the simulations with reduced radius and gravity have a smaller dayside cloud coverage (Figure 4b), possibly because they are hotter. In Section 4.1, we found that smaller radius leads to greater cloud cover, but when varying radius and gravity together the cloud cover reduces with decreasing radius. This implies that the effect of gravity outweighs the effect of radius on cloud cover. As a result, the inner edge of the habitable zone occurs at higher stellar irradiation when increasing radius and gravity together.
\section{Conclusions} \label{sec:conc}
In this work, we determined the inner edge of the habitable zone as we varied planetary radius, gravity, and surface pressure using the state-of-the-art GCM {\tt ExoCAM}. We compared our results to the 1D simulations of \cite{Kopparapu:2014}, and broke down the relative effects of radius and gravity on the inner edge of the habitable zone. From this work, we can draw the following key conclusions:
\begin{enumerate}
\item We find that the inner edge of the habitable zone moves toward higher stellar irradiation with increasing planetary mass, as found by \cite{Kopparapu:2014}. However, the inner edge occurs at a higher stellar irradiation in our simulations of planets orbiting M dwarf stars relative to the simulations of \cite{Kopparapu:2014} due to dayside clouds. 
\item Increasing planetary radius alone causes the inner edge of the habitable zone for planets orbiting M dwarf stars to move toward lower values of stellar irradiation. This is because increasing the radius leads to reduced cloud cover and reduced planetary albedo. The dynamical cause of this is a decrease in the Rossby radius of deformation relative to the planetary radius, causing the circulation to become asymmetric about the substellar point and reducing the cloud cover at the substellar point. 
\item Increasing surface gravity alone causes the inner edge of the habitable zone of M dwarf planets to move to higher values of stellar irradiation. This is because the $\mathrm{OLR}$ increases sharply with increasing gravity.
\item{ We find that the effects of gravity on the inner edge of the habitable zone outweigh the effects of planetary radius. As a result, increasing both radius and surface gravity together causes the inner edge of the habitable zone of M dwarf planets to move to higher values of stellar irradiation.}
\item {We find that the inner edge of the habitable zone for planets orbiting an M dwarf with an effective temperature of $3700 \ \mathrm{K}$ occurs at an orbital period of 41.72 days for $0.5R_\varoplus$ planets and an orbital period of 37.78 days for $1R_\varoplus$ planets. The \textit{TESS} mission will be able to find planets near the inner edge of the habitable zone orbiting M dwarf stars that can be characterized with \textit{JWST}. Our expectation that more massive terrestrial planets orbiting M dwarf stars will remain habitable at higher stellar irradiation can be tested with future observations of close-in terrestrial exoplanets.}
\end{enumerate}
\acknowledgements
We thank Eric Wolf for developing {\tt ExoCAM} and making it publicly available, and Daniel Koll for helpful discussions about tidally locked coordinates. We thank Ravi Kopparapu for a thorough review, which greatly improved the manuscript. Our work was completed with resources provided by the University of Chicago Research Computing Center and support from the NASA Astrobiology Institute's Virtual Planetary Laboratory, which is supported by NASA under cooperative agreement NNH05ZDA001C. T.D.K. acknowledges funding from the 51 Pegasi b Fellowship in Planetary Astronomy sponsored by the Heising-Simons Foundation.

\end{document}